\documentclass[conference]{IEEEtran}
\IEEEoverridecommandlockouts
\usepackage{cite}
\usepackage{amsmath,amssymb,amsfonts}
\usepackage{algorithmic}
\usepackage{graphicx}
\usepackage{textcomp}
\usepackage{xcolor}
\usepackage{epstopdf}
\usepackage{multicol}
\usepackage{diagbox}
\usepackage{multirow}
\usepackage{gensymb}
\def\BibTeX{{\rm B\kern-.05em{\sc i\kern-.025em b}\kern-.08em
    T\kern-.1667em\lower.7ex\hbox{E}\kern-.125emX}}
\begin{document}
\title{Channel Modeling and Signal Processing for Array-based Visible Light Communication System in Misalignment\\

}

\author{Jiaqi Wei, Chen Gong, Nuo Huang, and Zhengyuan Xu
	\thanks{ J. Wei, C. Gong, N. Huang and Z. Xu are with the Key Laboratory of Wireless-Optical  Communications, Chinese Academy of Sciences, School of Information Science and Technology, University of Science and Technology of China, Hefei 230027, China. Email: kellyway@mail.ustc.edu.cn, \{cgong821, huangnuo, xuzy\}@ustc.edu.cn..
	}}
\date{}

\maketitle

\begin{abstract}
This paper proposes an indoor visible light communication (VLC) system with multiple transmitters and receivers. Due to diffusivity of LED light beams, photodiode receive signals from many directions. We use one concave and one convex lens as optical antenna, and obtain the optimal lens structure by optimizing which corresponds to the minimum condition number of channel gain matrix. In this way the light emitted by different LED can be separated well from each other then minimize signal interference. However, interference increases in the case of system deviation, so we explore the system mobility. Then subsequent signal processing is carried out, including signal combining and successive interference cancellation (SIC). We combine the same signal received by different receivers to improve signal to interference noise ratio (SINR). And SIC can effectively restore interference and eliminate its impact. The simulation results show that channel capacity can be increased by more than 5 times and up to 20 times under the condition of receiver and transmitter alignment. In the case of movement, channel capacity can also be increased by about 4 times on average. Moreover, the mobile range of system is also significantly expanded.
\end{abstract}


\section{Introduction}
In the past decades, with continuous development of mobile communication technology, the speed of spectrum resource consumption is accelerating, and spectrum crisis is imminent\cite{background}. Researchers gradually pay more attention to visible light communication (VLC) that relies on visible light (VL) spectrum. It is a technology which uses LED to transmit data wirelessly by sending high-speed flashing signals. At the receiver, the light signal can be directly detected by a photodiode (PD) and restored to electrical signal. Then data transmission and lighting functions can be  simultaneously realized\cite{VLC}. VLC provides a feasible alternative to traditional communication methods and can be used as a supplement to the current RF communication. Thus it is an important field of wireless communication, and modern development of LED technology has greatly promoted the research of it. 

Since the feasibility of visible light communication was proposed\cite{model0}, a lot of work has been done in this field. Establishing channel model for visible light communication is an important basis. Some relevant research can be seen in \cite{model1}\cite{model2}\cite{model3}\cite{model4} and researchers are still improving it. Reference~\cite{raytracing} proposes a new channel modeling method based on ray tracing to describe the physical characteristics of environment more accurately. Since multiple LEDs are often used for illumination, VLC  is naturally combined with multiple input multiple output (MIMO) technology. In this way it can effectively achieve higher data rates while increasing brightness. Some articles like \cite{MIMO1}\cite{MIMO2}\cite{MIMO3}\cite{MIMO4}\cite{MIMO5} did prove the validity of this scheme. The diversity of MIMO system increases freedom of communication\cite{diversity}, and effectively combine signals at the receiver can further improve system performance\cite{MRC}. Therefore, more receivers can detect more useful signals and obtain higher performance gain. 

 In MIMO VLC system, imaging receiver structure is sometimes placed at the receiver to converge light beam, thus ensure a compact receiver array. For this kind of system, selecting a suitable optical antenna is an essential problem. Hemispherical lens\cite{hemisphical}, fisheye lens\cite{fisheye} and other different kinds of lenses\cite{convex}\cite{ACL} are often used to provide a wider view and increase optical gain. But lenses that can separate light beams and reduce inter-cross are also expected. Hence we will discuss about combination of concave and convex lens in this letter. Another tricky problem in MIMO VLC system is signal interference in displacement or misalignment state, which means different signals are probably received by the same receiver. To solve this problem, MIMO detection technology reconstructs and eliminates the interference at the receiver. And reference \cite{detection} conducts an extensive review of some representative methods. But still, we need method that takes into both high speed and mobility of communication systems into account.

In this paper, we establish an indoor visible light communication system with multiple transmitters and receivers. In order to suppress the interference among light beam, we use a lens group as optical antenna and talk about structure optimization. To better  apply to practice, we will explore the system mobility. Last we discuss the channel capacity gain brought by signal processing method we use. The main point of this paper is to propose the MIMO VLC system build process, including antenna optimization and signal processing, which effectively improves the channel capacity and mobility of the system.

The rest of this paper is organized as follows: Section \ref{System Model} shows the indoor VLC system we established.  Section \ref{Signal Processing} describes how to determine the optical antenna parameters in this VLC system and introduces the subsequent signal processing method in detail. Simulation results and evaluation are shown in Section \ref{Result Analysis}. At last, the paper will be concluded in Section \ref{Conclusion}.

\section{System Model}
\label{System Model}

We consider a typical array-based VLC system including an LED array, a PD array and two lenses for concentrating light. Both then LED array and PD array consist of $N \times N$ elements; and the two lenses include a convex lens and a concave lens, which can narrow the LED beam into parallel light and project the light from each LED to its corresponding PD component. 

\begin{figure}[htbp]
	\centerline{\includegraphics[height=8cm,width=5.5cm]{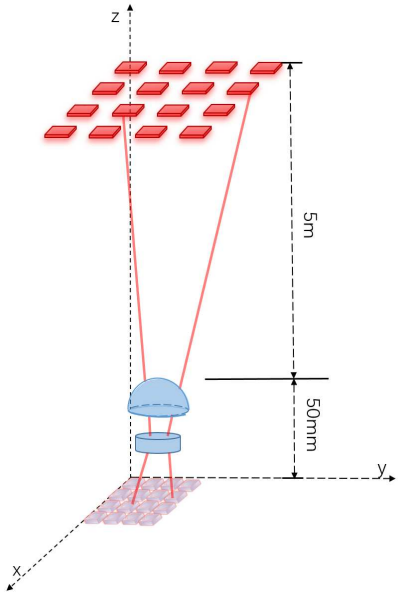}}
	\caption{Configuration of the array-based VLC system under consideration.}
	\label{fig1}
\end{figure}

An example is shown in Fig.~\ref{fig1}, where the LED array and PD array are both composed of $4 \times 4$ components. The distance between transmitter plane and receiver plane is set to be $5050 mm$, which can meet the requirements of most daily scenes. The key parameters are shown in TABLE~\ref{table1}.

\begin{table}[htbp]
	\caption{key parameters of the array-based visble light communication system}
	\begin{center}
		\renewcommand\arraystretch{1.5}
		\begin{tabular}{|c|c|c|c|}
			\hline
			\textbf{} & \textbf{\textit{LED}}& \textbf{\textit{Lens}}& \textbf{\textit{PD}} \\
			\hline
			\textbf{Size/mm}&$10\times10$ &$\phi15,\phi10$ & $0.6\times0.6$  \\
			\hline
			\textbf{Quantity}&$4\times4$ &2 & $4\times4$  \\
			\hline
			\textbf{Element gap/mm}&10 &20 & 0.1  \\
			\hline
			\textbf{Height/mm}&5050 &50,30 & 0  \\
			\hline
		\end{tabular}
		\label{table1}
	\end{center}
\end{table}

Assume that the LED has cosine radiation intensity given by
\begin{equation}
{I(\phi)}\approx {I_{0}(cos\phi)^{C_n} },
\label{eq11}
\end{equation}
where $\phi$ is the angle between the light and the normal line; $I_0$ is the radiation intensity along the normal direction. The exponent $C_n$ is greater than or equal to 1, but does not need to be an integer. Larger $C_n$ enables narrower intensity distribution of light source. For example, we can set $C_n$ to be 10, which means that the half-power angle is approximately $21\degree$.

In this work, we assume that the reflection component is significantly weaker than the line-of-sight (LOS) component, and thus the LOS component dominates. The communication system can be modeled as a MIMO communication system with $N_t$ transmitters and $N_r$ receivers. The noise can be modeled as the additive white Gaussian noise.


\section{Lens Optimization and Signal Processing}
\label{Signal Processing}
The two optical lenses need to be specially designed such that the light from each LED can be projected to the corresponding PD. Then, the diagonal components of the corresponding channel gain matrix dominate. The lens optimization and the related signal processing method will be detailed in this section.

\subsection{Lens Optimization}\label{AA}
In the proposed system, we use two aspheric lenses, one convex and one concave, such that the signals from different LEDs can be separated. The surfaces of both lenses are paraboloid and the expressions can be written as

\begin{equation}
z=\frac{cr^2}{1+\sqrt{1-(1+k)c^2r^2}}+\alpha r^2+\beta r^4+\cdots,
\label{eq2}
\end{equation}
where $z$ is the 3-D surface of lens, and $r$ represents the distance from the lens edge to central axis. For simplicity, we consider coefficient of quadratic term $\alpha$, and set the radius of curvature $c$ and conic constant $k$ for each lens to be zero\cite{lensdesign}.

We adopt Zemax, a comprehensive optical design and simulation software, for the lens optimization. Given parameters like geometric position and refractive index of the object, the direction and intensity of light can be determined. We simulated the array-based VLC system model using Zemax software and figure out how light travels. Then, we can readily obtain the channel gain matrix through extension program with MATLAB. 


\begin{table}[htbp]
	\caption{ The optimized parameters of two lenses}
	\begin{center}
		\renewcommand\arraystretch{1.5}
		\begin{tabular}{|c|c|c|c|}
			\hline
			\textbf{} & \textbf{\textit{$\alpha_1$}}& \textbf{\textit{$\alpha_2$}}& \textbf{\textit{thickness/mm}} \\
			\hline
			\textbf{Convex Lens}&0.036 &0.007& 6.875 \\
			\hline
			\textbf{Concave Lens}&-0.08 &0.05 &2 \\
			\hline
		\end{tabular}
		\label{table2}
	\end{center}
\end{table}

\begin{figure}[htbp]
	\centerline{\includegraphics[height=5cm,width=8cm]{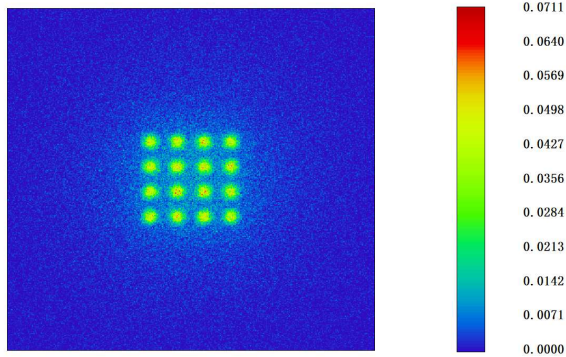}}
	\caption{Spot pattern on the detector plane using a $10mm\times10mm$ detector.}
	\label{fig5}
\end{figure}

We optimize the structure of the two lens to minimize the condition number of channel gain matrix. The specific parameters to be optimized are $\alpha^1_1,\alpha^1_2,\alpha^2_1,\alpha^2_2,$ corresponding to the front and rear surfaces of two lenses, where superscript 1 of  $\alpha$ denotes convex lens and superscript 2 denotes concave lens, and subscript 1 indicates front surface and subscript 2 indicates the back surface. Through iterative optimization, the optimal parameters of the above mentioned example can be obtained, as shown in TABLE~\ref{table2}. The minimum condition number 1.6622, and the spot pattern on detector plane is shown in Fig.~\ref{fig5}.

\begin{figure*}[htbp]
	\small{
		\[
		\setlength{\arraycolsep}{1pt}
		\boldsymbol{H} =\left( \begin{array}{cccccccccccccccc}  
		445.88 & 27.71 & 11.90 & 6.40 & 27.69 & 17.94 & 9.75 & 6.28 & 11.10 & 9.75 & 7.57 & 5.37 & 6.37 & 5.93 & 5.01 & 3.91 \\
		31.94 & 450.20 & 29.00 & 11.61 & 18.14 & 28.00 & 17.75 & 10.06 & 10.18 & 11.40 & 9.67 & 7.35 & 5.98 & 6.70 & 6.20 & 5.54 \\
		12.05 & 29.67 & 451.72 & 29.95 & 9.80 & 17.46 & 28.14 & 18.46 & 7.56 & 10.24 & 10.99 & 9.71 & 4.75 & 6.37 & 6.55 & 6.07 \\
		5.95 & 11.07 & 27.88 & 443.46 & 6.47 & 9.53 & 17.06 & 27.42 & 5.21 & 7.09 & 9.23 & 11.39 & 3.79 & 5.38 & 5.94 & 6.81 \\
		31.40 & 18.29 & 9.97 & 6.32 & 450.96 & 27.82 & 10.62 & 6.54 & 29.00 & 17.49 & 10.35 & 6.28 & 11.61 & 9.41 & 7.07 & 4.91\\ 
		19.04 & 32.04 & 18.73 & 9.95 & 31.03 & 452.33 & 29.08 & 11.18 & 18.54 & 29.31 & 18.15 & 10.18 & 9.64 & 11.05 & 9.81 & 7.04\\ 
		9.56 & 19.28 & 31.43 & 19.45 & 11.98 & 29.42 & 452.16 & 31.05 & 10.28 & 18.43 & 28.87 & 18.69 & 7.09 & 10.14 & 11.35 & 9.73 \\
		5.88 & 10.16 & 18.47 & 29.93 & 6.55 & 10.96 & 27.64 & 446.16 & 6.32 & 9.39 & 17.47 & 29.61 & 5.42 & 7.49 & 9.21 & 11.18 \\
		11.37 & 9.75 & 6.90 & 4.76 & 28.46 & 18.11 & 9.66 & 6.04 & 453.32 & 28.12 & 11.88 & 6.35 & 30.28 & 18.59 & 9.76 & 5.85 \\
		10.39 & 11.99 & 10.33 & 6.97 & 18.61 & 29.72 & 18.08 & 9.87 & 30.34 & 457.93 & 29.33 & 11.70 & 18.94 & 32.27 & 18.73 & 9.66\\ 
		7.14 & 9.81 & 11.29 & 9.83 & 9.63 & 18.36 & 29.46 & 18.71 & 11.31 & 30.12 & 456.41 & 31.12 & 9.62 & 18.77 & 31.58 & 19.15 \\
		5.20 & 7.07 & 9.66 & 11.89 & 6.03 & 9.50 & 17.86 & 29.22 & 6.56 & 11.62 & 27.06 & 450.99 & 6.00 & 9.65 & 17.78 & 32.14 \\
		6.68 & 6.03 & 5.12 & 4.22 & 11.45 & 10.15 & 7.41 & 4.96 & 28.72 & 17.53 & 9.39 & 6.25 & 447.29 & 27.78 & 10.68 & 6.96 \\
		5.77 & 6.24 & 5.77 & 5.39 & 9.73 & 11.63 & 9.86 & 6.75 & 18.46 & 27.19 & 17.89 & 9.88 & 31.15 & 455.03 & 28.76 & 11.40 \\
		5.09 & 6.12 & 6.70 & 5.92 & 6.74 & 9.56 & 11.35 & 9.73 & 9.79 & 17.27 & 27.15 & 17.51 & 10.93 & 28.74 & 447.49 & 30.92 \\
		4.10 & 5.45 & 5.95 & 6.40 & 5.35 & 6.73 & 9.47 & 10.99 & 6.13 & 9.61 & 17.48 & 27.50 & 6.64 & 11.27 & 28.23 & 444.46 
		\end{array}
		\right)  \times10^{-7}               
		\]}
	\caption{Channel gain matrix of aligned transmitter and receiver}
	\label{channel}
\end{figure*}

From Fig.~\ref{fig5}, it can be seen that the intervals between different light spots are easy to distinguish, which implies that the light beam can be well separated and interference can be suppressed. Moreover, the size of each spot is about $0.5mm\times0.5mm$, which is about 20 times smaller than that of LED ($10mm\times10mm$), which proves that the optical antenna has good concentrating ability. The corresponding channel gain matrix is shown in Fig.~\ref{channel}, where the diagonal elements are significantly larger than the off-diagonal ones.

\begin{figure*}[htbp]
	\centerline{\includegraphics[height=5cm,width=14cm]{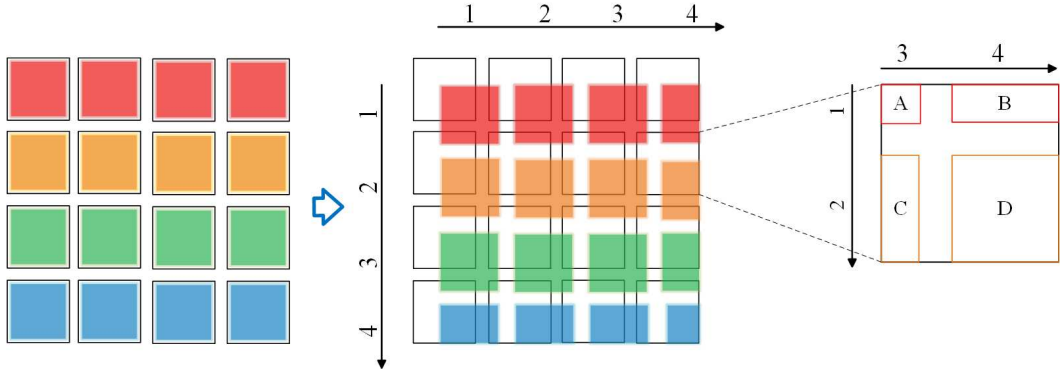}}
	\caption{Spot pattern of mobile system.}
	\label{spot}
\end{figure*}

Consider the scenario where the transmitter and receiver are not in perfect alignment. The spatial distribution of light from the LEDs deviates from the position of the PD array, resulting in one light pot covering multiple adjacent PDs and one PD covered by the pots from multiple adjacent LEDs, as shown in Fig.~\ref{spot}. There are sixteen rectangular detectors in this figure, and the colored rectangles indicate the LED spots on detector. Clearly, spots move with the senders, and reach the junction of two or four detectors. Meanwhile, part of one spot has exceeded the detection area.

It is seen that since each receiver may be covered by the lights pots from multiple LEDs, the communication can be modeled by an interference channel. However, due to the specific structure of the interference channel, the transmitter information can be extracted by successive interference cancellation (SIC). The detailed process is elaborated in the subsequent subsections.

\subsection{Signal Processing for Array-based System}

\begin{figure}[htbp]
	\centerline{\includegraphics[height=3.5cm,width=6.5cm]{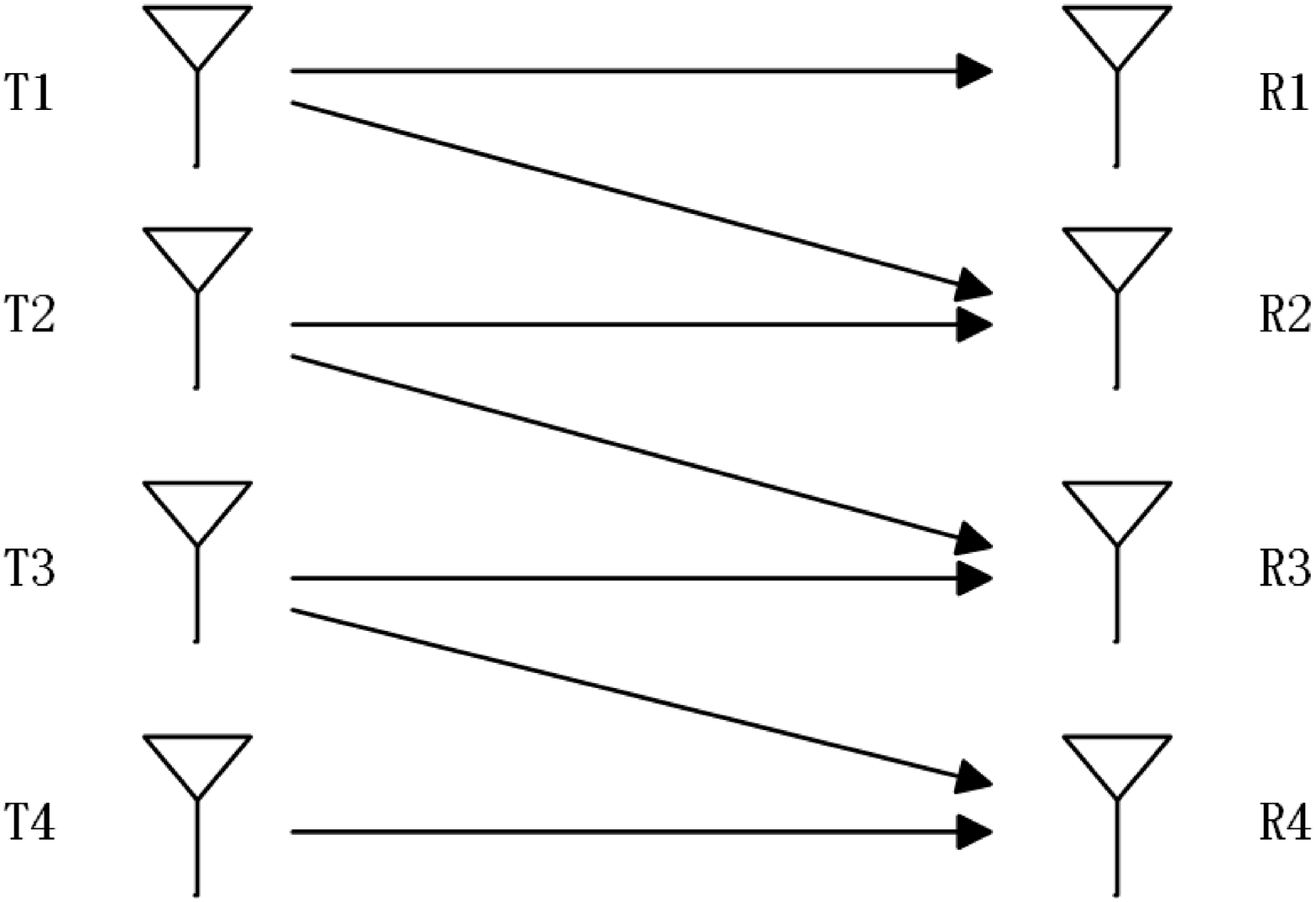}}
	\caption{Channel model of mobile system.}
	\label{zpng}
\end{figure}

We already know the channel structure, and the channel model of Fig.~\ref{spot} can be represented as shown in Fig.~\ref{zpng}, where T1, T2, T3 and T4 are transmitters, R1, R2, R3 and R4 are receivers. This kind of channel can be seen as a superposition of Z-channel. Z-channel is a two input two output channel model, whereby there are three transmission paths. Two senders respectively connect to the corresponding receiver and the transmission from one of the senders introduces interference to the non-intended receiver, while the other receiver is free from interference\cite{zchannel}.

Since one receiver is interference-free, we can detect this signal first, then get the original signal from transmitter. After that, it is easy to know the interference information. This is undoubtedly the idea of SIC\cite{SIC}. In Fig.~\ref{spot}, we can clearly see that the detector in the top left corner receives the signal without interference, so the demodulation will start from this one. Once the first signal is known, it is easy to get the interference caused by it. Next we can demodulate signals received by the two detectors to fist one's right and below. In this way, we mark the numbers $1,2,3,4$ around the spots, which represents the order of demodulating. 

As for the signals that fall on the middle detectors, we particularly discuss the order of recovering them. Take an example shown on the right side of Fig.\ref{spot}, the detector received four different signals,denoted as $A, B, C$ and $D$. Vertically, we first demodulate signals $A$ and $C$, which are labeled as 1 and 2, and next one is $B$. The signal $D$ in the lower right corner will be restored finally after all interference is eliminated.

In a word, the process of SIC mainly includes four parts: selecting the signal with maximum power, signal decision, signal reconstruction and interference cancellation. Generally, the signal with high power has a low bit error rate, so successively processing can reconstruct and eliminate interference as accurately as possible. 

However, in MIMO system signal always has multiple transmission paths. This kind of diversity makes it possible for signals to have different transmission quality. So we are able to combine those received signals for higher SINR before interference cancellation. 

Since the light from the same LED falls on 4 detectors at most, we select 4 PDs which received the strongest light power from each sender. After that, we adopt the idea of maximal ratio combining (MRC) to sum the 4 received signals. We set different weights for different links, and the coefficient multiplied by received signal is proportional to the DC gain, so the weight vector can be expressed as
\begin{equation}
\boldsymbol w_j = \frac{\begin{pmatrix}\begin{smallmatrix}
 0, &\dots, &h_{mj_1,j},&0, &\dots, &h_{mj_2,j},&0,& \dots, & h_{mj_3,j}, &0, & \dots, & h_{mj_4,j}, & 0, & \dots
\end{smallmatrix}
\end{pmatrix}}{h^2_{mj_1,j}+h^2_{mj_2,j}+h^2_{mj_3,j}+h^2_{mj_4,j}},
\end{equation}
where $\boldsymbol w_j$ is corresponding to $j$-th transmitting signal, and $h_{mj,j}$ is the elemant of channel gain matrix. There are four non-zero elements in weight vector, meaning we combine four received signals into one.

The combined signal is
\begin{equation}
\boldsymbol y^{co} =\boldsymbol w \boldsymbol y,
\end{equation}
where $\boldsymbol{y}$ is the original received signal.

Integrate this with SIC method, the $j$-th combined signal turns to be

\begin{equation}
\boldsymbol y^{co}_{j} =\boldsymbol w_j \boldsymbol y_{-(j-1)} = \boldsymbol w_j \boldsymbol h_jx_j+ \boldsymbol w_j\sum^{N_t}_{i=j+1}{\boldsymbol h_ix_i+ \boldsymbol w_j \boldsymbol n},
\end{equation}
where $\boldsymbol y_{-(j-1)}$ denotes the $j$-th received signal that has eliminated interference caused by $1st$ to $(j-1)$-th signal,  $\boldsymbol n$ is the white Gaussian noise and its variance is $\sigma_n$.

And SINR is
\begin{equation}
SINR_{j}= \frac{P_{S_j}}{P_{I_j}+P_{N_j}}=\frac{(\boldsymbol w_j \boldsymbol h_j)^2}{\sum^{N_t}_{i=j+1}(\boldsymbol w_j\boldsymbol h_i)^2+\vert\vert\boldsymbol w_j \boldsymbol n\vert\vert^2},
\label{SINR}
\end{equation}
where $P_{S_j}$ is signal power, $P_{I_j}$ is total interference power, and $P_{N_j}$ is noise power. In this way, the channel capacity ($bps/Hz$ measurement) can be expressed as
\begin{equation}
C_j=log(1+SINR_j),
\label{C}
\end{equation}
where $C_j$ denotes channel capacity of the $j$-th sub-channel.

\subsection{Signal Processing for the Proposed System}
 There are 16 LEDs and 16 PDs in the previously established system. Due to the fixed size of light spot, the transmitter has a limited moving range. And we prefer system with greater range of motion, naturally adding more PDs on the receiving plane is a feasible way. We changed the previous 4$\times$4 PD array to 8$\times$8, while PD size and interval keeping the same. As LEDs move, more PDs will be able to receive signals, thus ensuring a larger range of movement.

We already know the channel state information, thus we know which detectors receive the strongest signal. Before signal processing, we should sort the signals by power. When detecting the first signal, the received signals can be written as
\begin{equation}
\boldsymbol y = \boldsymbol h_1x_1+\sum_{i=2}^{16}\boldsymbol h_{i}x_{i}+ \boldsymbol n.
\end{equation}

Assume that we combine the first, second, third and fourth received signals into one, the weight vector is
\begin{equation}
\boldsymbol w_1 = \frac{\begin{pmatrix}
h_{11},&h_{21},& h_{31},& h_{41}, & 0, & \dots,&0
	\end{pmatrix}}{h^2_{11}+h^2_{21}+h^2_{31}+h^2_{41}}.
\end{equation}
So the combined signal could be expressed as 
\begin{equation}
\boldsymbol y^{co}_{1} =\boldsymbol w_1 \boldsymbol y =  \boldsymbol w_1\sum^{16}_{i=1}{\boldsymbol h_ix_i+ \boldsymbol w_1 \boldsymbol n},
\end{equation}
and SINR turns to be 
\begin{equation}
SINR_{1}=\frac{(\boldsymbol w_1 \boldsymbol h_1)^2}{\sum^{16}_{i=2}(\boldsymbol w_1\boldsymbol h_i)^2+\vert\vert\boldsymbol w_1 \boldsymbol n\vert\vert^2}.
\label{SINR}
\end{equation}

Original signal $x_1$ can be recovered through decision. Thus we reconstruct the interference introduced by $x_1$ and conduct interference cancellation. The same processing will be performed on all signals in turn to ensure that all interference is eliminated. Again we use equations (\ref{SINR}) and (\ref{C}) to calculate channel capacity.

\section{Result Analysis}
\label{Result Analysis}
\setcounter{subsection}{0}
This section presents the numerical simulation result of signal processing. We compare the change in channel capacity when transmitter and receiver translate or rotate respectively. Then we increase to 64 PDs to see the enhancement of system mobility. 

\subsection{Moving Range}
\begin{figure}[htbp]
\centerline{\includegraphics[height=4.5cm,width=8cm]{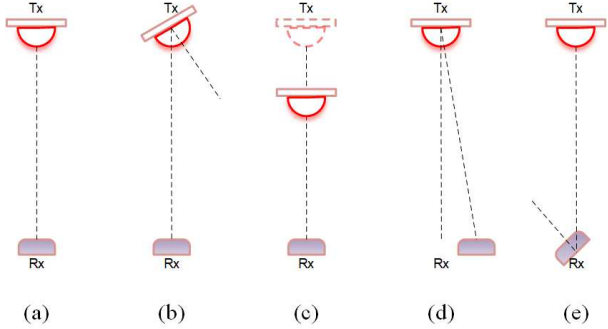}}
\caption{Relative positions of transmitter and receiver when (a)center aligned (b)transmitter rotates (c)transmitter vertically translates (d)receiver horizontally moves (e)receiver rotates.}
\label{fig4}
\end{figure}

In practical applications, system may be inclined or in offset when placed, such as Fig.~\ref{fig4}, which makes the center of transmitter and receiver cannot be perfectly aligned. Therefore, we first explore how channel characteristics change when system move.

\begin{figure}[htbp]
\centerline{\includegraphics[height=4.5cm,width=8cm]{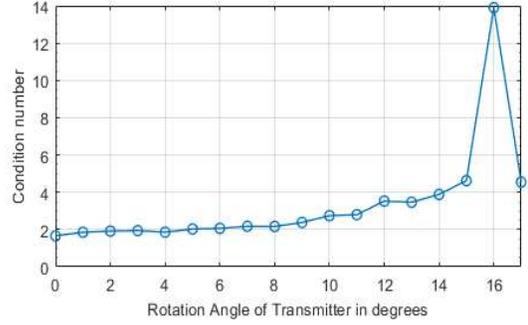}}
\caption{Variation of condition number with transmitter rotation.}
\label{fig2}
\end{figure}

\begin{figure}[htbp]
\centerline{\includegraphics[height=5cm,width=8cm]{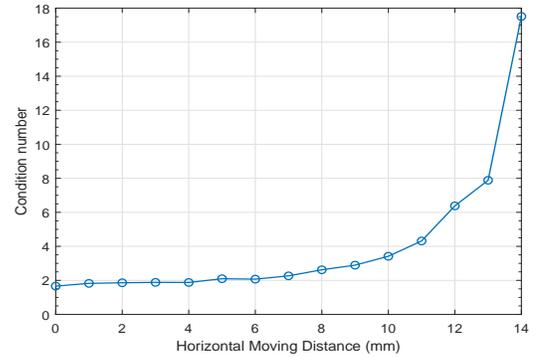}}
\caption{Variation of condition number with receiver horizontal translation.}
\label{fig3}
\end{figure}

\begin{figure}[htbp]
	\centerline{\includegraphics[height=4cm,width=6cm]{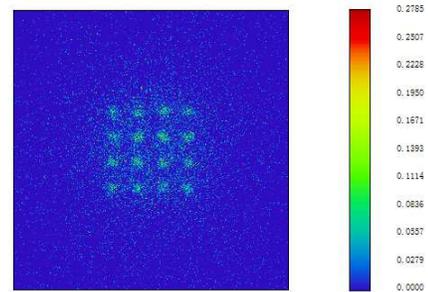}}
	\caption{Spot pattern on detector plane after vertically moving $1m$. }
	\label{vertical}
\end{figure}

\begin{table*}[htbp]
	\caption{Channel capacity of rotating receiver}
	\begin{center}
		\renewcommand\arraystretch{1.2} 
		\setlength{\tabcolsep}{4.5mm}
		\begin{tabular}{c|c|ccccccccc}
			\hline
			\multicolumn{2}{c|}{\diagbox{channel capacity}{signal number}}& 1 & 2 & 3 & 4 & 5 & 6 & 7 & 8  \\
			\hline
			\multirow{4}{*}{Rotate $0\degree$}& SC+SIC &4.789  & 5.678  & 12.972  & 10.700  & 5.443  & 3.918  & 4.525  & 5.489 \\
			&SIC only &2.089  & 2.608  & 4.356  & 2.202  & 2.043  & 2.001  & 2.001  & 2.638 \\
			& SC only&0.965  & 0.964  & 0.977  & 0.977  & 0.964  & 0.951  & 0.951  & 0.964  \\
			& No processing&0.986  & 0.986  & 0.989  & 0.989  & 0.986  & 0.982  & 0.982  & 0.9869 \\
			\hline
			\multirow{4}{*}{Rotate $0.5\degree$}& SC+SIC &2.816 & 2.477 & 2.541 & 9.908 & 4.586 & 2.091 & 1.912 & 2.027\\
			&SIC only &0.699 & 0.007 & 0.269 & 2.083 & 0.723 & 0.154 & 0.003 & 0.114\\
			& SC only&0.779 & 0.743 & 0.85 & 0.891 & 0.849 & 0.732 & 0.7 & 0.811 \\
			& No processing&0.823  & 0.682  & 0.976  & 0.982  & 0.975  & 0.804  & 0.667  & 0.966  \\
			\hline
			\multirow{4}{*}{Rotate $1\degree$}& SC+SIC &3.353  & 5.521  & 6.402  & 7.209  & 0.697  & 4.578  & 4.595  & 4.984  \\
			&SIC only &1.638  & 0.255  & 0.033  & 0.744  & 0.446  & 0.263  & 0.290  & 0.165  \\
			& SC only&0.009  & 0.965  & 0.977  & 0.965  & 0.007  & 0.952  & 0.951  & 0.965 \\
			& No processing&0.004  & 0.986  & 0.989  & 0.986  & 0.004  & 0.983  & 0.982  & 0.986 \\
			\hline
			\end{tabular}
		
\begin{tabular}{ccccccccccc}
			\\
			 \\
\end{tabular}

 \begin{tabular}{c|c|ccccccccc}
 \hline
			\multicolumn{2}{c|}{\diagbox{channel capacity}{signal number}} & 9 & 10 & 11 & 12 & 13 & 14 & 15 & 16 \\
			\hline
			\multirow{4}{*}{Rotate $0\degree$}& SC+SIC & 4.669  & 4.393  & 3.951  & 4.716  & 4.782  & 5.758  & 10.447  & 20.917 \\
			&SIC only & 2.001  & 2.007  & 2.007  & 2.006  & 2.358  & 2.076  & 2.001  & 10.812 \\
			& SC only& 0.965  & 0.951  & 0.951  & 0.965  & 0.965  & 0.964  & 0.978  & 0.977 \\
			& No processing& 0.986  & 0.982  & 0.982  & 0.986  & 0.985  & 0.986  & 0.989  & 0.989 \\
			\hline
			\multirow{4}{*}{Rotate $0.5\degree$}& SC+SIC & 3.691 & 2.145 & 1.874 & 2.185 & 2.829 & 2.482 & 2.535 & 19.706\\
			&SIC only & 0.399 & 0.268 & 0.017 & 0.02 & 0.23 & 0.528 & 0.677 & 8.224\\
			& SC only & 0.848 & 0.727 & 0.702 & 0.81 & 0.776 & 0.743 & 0.851 & 0.889\\
			& No processing & 0.974  & 0.798  & 0.665  & 0.966  & 0.822  & 0.677  & 0.976  & 0.982 \\
			\hline
			\multirow{4}{*}{Rotate $1\degree$} &SC+SIC & 1.038 & 5.221 & 4.143  & 4.409  & 5.477  & 13.676  & 5.433  & 7.191 \\
			&SIC only & 0.574  & 0.308  & 0.267  & 0.257  & 0.406  & 7.498  & 6.864  & 0.315 \\
			& SC only & 0.007  & 0.952  & 0.952  & 0.966  & 0.009  & 0.965  & 0.978  & 0.964 \\
			& No processing & 0.004  & 0.982  & 0.983  & 0.986  & 0.004  & 0.986  & 0.990  & 0.986 \\
			\hline
		\end{tabular}
		\label{simulation11}
	\end{center}
\end{table*}

\begin{table*}[htbp]
	\caption{Channel capacity of horizontally moving receiver}
	\begin{center}
		\renewcommand\arraystretch{1.2}  
		\setlength{\tabcolsep}{4.5mm}
		\begin{tabular}{c|c|cccccccc}
			\hline
			\multicolumn{2}{c|}{\diagbox{channel capacity}{signal number}}& 1 & 2 & 3 & 4 & 5 & 6 & 7 & 8\\
			\hline
			\multirow{4}{*}{Move $50mm$}& SC+SIC &5.804  & 4.025  & 7.526  & 11.526  & 4.684  & 3.908  & 3.081  & 4.443 \\
			&SIC only &1.931  & 1.130  & 1.965  & 3.039  & 1.443  & 1.082  & 1.068  & 1.001  \\
			& SC only&0.947  & 0.930  & 0.944  & 0.964  & 0.951  & 0.930  & 0.912  & 0.920 \\
			& No processing&0.980  & 0.973  & 0.965  & 0.990  & 0.988  & 0.976  & 0.968  & 0.958  \\
			\hline
			\multirow{4}{*}{Move $100mm$}& SC+SIC &5.08 & 2.55 & 2.863 & 2.785 & 0.609 & 2.47 & 2.065 & 2.563\\
			&SIC only &0.436  & 1.576  & 0.412  & 0.243  & 0.466  & 0.274  & 0.205  & 0.198\\
			& SC only&0.014  & 0.773  & 0.885  & 0.507  & 0.011  & 0.480  & 0.733  & 0.851  \\
			& No processing&0.006  & 0.798  & 0.981  & 0.594  & 0.006  & 0.590  & 0.791  & 0.974  \\
			\hline
			\multirow{4}{*}{Move $150mm$}& SC+SIC &0.742  & 1.006  & 6.782  & 9.929  & 0.495  & 0.368  & 0.436  & 4.416 \\
			&SIC only &0.423  & 0.243  & 0.156  & 6.330  & 0.494  & 0.248  & 0.170  & 0.144 \\
			& SC only&0.003  & 0.010  & 0.001  & 0.979  & 0.001  & 0.002  & 0.007  & 0.966 \\
			& No processing&0.004  & 0.001  & 0.000  & 0.991  & 0.004  & 0.004  & 0.004  & 0.988 \\
			
			\hline
		\end{tabular}
		
		\begin{tabular}{ccccccccccc}
			\\
			\\
		\end{tabular}
		
		\begin{tabular}{c|c|ccccccccc}
			\hline
			\multicolumn{2}{c|}{\diagbox{channel capacity}{signal number}}& 9 & 10 & 11 & 12 & 13 & 14 & 15 & 16 \\
			\hline
			\multirow{4}{*}{Move $50mm$}& SC+SIC & 5.616  & 4.497  & 3.953  & 3.351  & 5.794  & 3.879  & 7.691  & 20.715 \\
			&SIC only  & 1.520  & 1.210  & 1.007  & 1.123  & 1.250  & 1.188  & 1.784  & 9.094 \\
			& SC only & 0.950  & 0.931  & 0.912  & 0.923  & 0.948  & 0.929  & 0.943  & 0.965 \\
			& No processing& 0.987  & 0.975  & 0.968  & 0.959  & 0.981  & 0.973  & 0.964  & 0.991 \\
			\hline
			\multirow{4}{*}{Move $100mm$} & SC+SIC & 0.856 & 3.108 & 2.125 & 2.319 & 5.178 & 2.543 & 2.831 & 12.563\\
			&SIC only  & 0.517  & 0.303  & 0.320  & 0.243  & 0.061  & 0.611  & 0.942  & 6.820 \\
			& SC only& 0.011  & 0.479  & 0.730  & 0.848  & 0.015  & 0.772  & 0.505  & 0.883 \\
			& No processing& 0.006  & 0.587  & 0.785  & 0.974  & 0.006  & 0.793  & 0.590  & 0.980 \\
			\hline
			\multirow{4}{*}{Move $150mm$}& SC+SIC & 0.731  & 0.476  & 0.565  & 5.083  & 0.936  & 1.012  & 1.409  & 6.856 \\
			&SIC only  & 0.564  & 0.299  & 0.213  & 0.268  & 0.423  & 0.443  & 0.696  & 0.642 \\
			& SC only  & 0.001  & 0.002  & 0.007  & 0.966  & 0.003  & 0.010  & 0.001  & 0.979 \\
			& No processing & 0.001  & 0.001  & 0.001  & 0.987  & 0.000  & 0.000  & 0.000  & 0.991 \\
			
			\hline
		\end{tabular}
		\label{simulation21}
	\end{center}
\end{table*}

\begin{figure}[htbp]
	\centerline{\includegraphics[height=6cm,width=8.5cm]{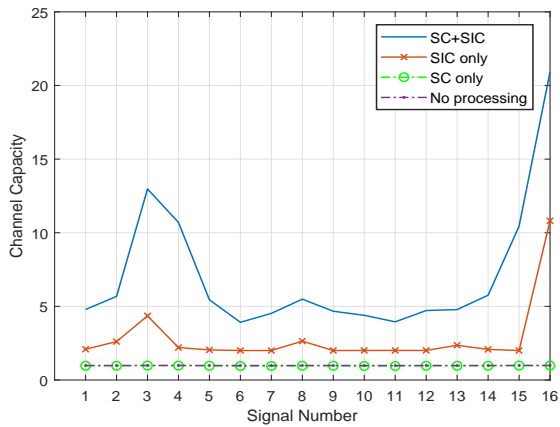}}
	\caption{Comparison of channel capacity using different signal processing methods without movement.}
	\label{capacity1}
\end{figure}

\begin{figure}[htbp]
	\centerline{\includegraphics[height=6cm,width=9cm]{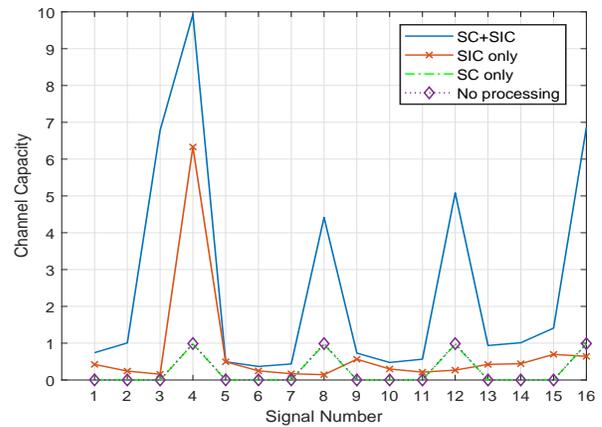}}
	\caption{Comparison of channel capacity using different signal processing methods with horizontal movement of $150 mm$.}
	\label{capacity2}
\end{figure}

Here we assume that the LED array is encapsulated, i.e., the relative position between components will not change as array moves. And the same assumption for PD array. Still, we choose condition number as representative of channel performance. We get the results of condition number in situation of system rotates and translates, shown in Fig.~\ref{fig2} and Fig.~\ref{fig3}. We can see that when receiver is perfectly aligned (rotate 0$\degree$ in Fig.~\ref{fig2}), the smallest condition number representing the best channel performance is obtained. As receiver moves away, the condition number increases gradually, showing the transmission performance is getting worse.

Only when channel performance is in a certain range can signal be transmitted with high quality. So we consider the condition numbers less than 10 indicating good transmission condition. Hence the movable range can be determined. In Fig.~\ref{fig2} condition number increases gradually, but suddenly drops at $x=17\degree$. That's because channel condition deteriorates as the transmitter rotates. And PD can not receive optical signal starting from a certain position, so all sub-channels have the same poor performance, resulting in the reduction of condition number.

In Fig.~\ref{fig3} the receiver can move about 13mm horizontally while maintaining good signal transmission condition. But in vertical direction, the receiver (or transmitter) can move about $\pm1m$. Fig.~\ref{vertical} shows the spot.

Notwithstanding, the receiver has a small rotation range of less than 1 degree because lens rotation can change the light path seriously.

\subsection{Signal Processing Result}  
Next we will discuss the results of signal processing and start with the situation of receiver moving.

We simulate the numerical results of channel capacity using signal combining and SIC methods (SC+SIC) under different mobile situations. As comparison, we also calculate the channel capacity in the case of signal combining without SIC step (SC only) and SIC without signal combining step (SIC only) respectively.
 
We study the situation of receiver rotation. Since sender-side rotation is equivalent to receiver-side rotation along with translation, the rotation angle range of sender is larger than that of receiver. The simulation results in TABLE~\ref{simulation11} show a significant improvement in channel capacity when using SC+SIC, meaning that each channel can transfer information at a higher rate. And rotation angle increasing to 1 degree is feasible now.

The receiver has a limited moving distance in the horizontal direction because of small receiving area. But in TABLE~\ref{simulation21}, adding PDs and signal processing could effectively improve channel capacity and mobility at the same time. It is obvious the channel capacity of each subchannel has increased by more than  four times. Moreover, receiver is able to move about $150 mm$, which is $137 mm$ longer than before.

As for offsetting transmitter, performing SC and SIC also improves system performance. But when sender moves vertically, the communication performance is almost unchanged, so the performance gain brought by signal processing is not obvious.
 
Next we analyze the simulation results in more detail.  We focus on the situation of alignment shown in Fig.~\ref{capacity1}. At this time the 16-channel signals correspond to PD one by one, i.e., the interference is minimal. So we can see sub-channels have the same original channel capacity. But after signal processing, the maximum capacity of sub-channel can be up to 20, which is about 20 times what it was before.

Then we discuss the channel capacity under offset. Taking horizontal movement of $150 mm$ as an example, we can see changes in channel capacity when using different signal processing methods is shown in Fig.~\ref{capacity2}. After SC and SIC steps, the highest channel capacity is obtained, corresponding to the top curve in the graph. There exist four peaks because only four PDs located at the edge are able to receive signals in such a situation. Independent SIC processing has a small effect on channel capacity since its performance can be affected by signal detection. When signal power is pretty low, incorrect judgment of signal will cause progressive performance degradation. However, signal combining has no contribution to improve SNR as other PDs can hardly receive signals.

In a word, by increasing the number of PD, the moving range of system is obviously increased, so system can move further. In addition to that, signal combining can effectively improve SINR and SIC can eliminate interference step by step. Whether LED and PD are aligned or in offset, using SC with SIC method can both significantly improve the achievable rate of each sub-channel.

\section{Conclusion}
\label{Conclusion}
In this paper, we propose a long-distance MIMO visible light communication system and corresponding signal processing method. After optimizing, one convex and one concave lenses structure are determined. Using them as optical antennas, spots of $4\times4$ LEDs are clearly separated on the receiving plane with obvious contour shape at a distance of 5m. This means each PD can receive the signal from corresponding LED with only little interference. But in practice, there are often scenarios where the transmitter and receiver cannot be perfectly aligned. So we study the influence on transmission performance of system caused by translation and rotation. Besides, we explore the system mobility.

Moreover, we use the signal combining and successive interference cancellation method to suppress the reduction of channel capacity caused by movement. Adding PDs to $8\times8$, the mobility of the system is significantly improved. But signals from one LED may be received by multiple PDs, so we weightedly merge the same transmitting signal and perform successive interference cancellation on the combined signal. As a result, channel capacity of each sub-channel is increased by more than four times after signal processing, thus higher data rates can be achieved.

\end{document}